\documentstyle[12pt,leqno]{article}
\textheight9in \def\DATE{June 4, 1996}
\newtheorem{theorem}{Theorem}[section]

\newtheorem{observation}[theorem]{Observation}
\newtheorem{odstavec}[theorem]{}

\newtheorem{lemma}[theorem]{Lemma}

\newtheorem{proposition}[theorem]{Proposition}

\catcode`\@=11

\def\ps@myheadings{\let\@mkboth\@gobbletwo
\def\@oddhead{\ifnum\count0=1 \hfill\else
\rightmark \hfil \rm\thepage\fi}%
\def\@oddfoot{\ifnum\count0=1 \hfill \rm 1 \hfill \else
\hfill\fi}
\def\@evenhead%
{\rm\leftmark\hfil\rm\thepage}%
\def\@evenfoot{}\def\sectionmark##1{}
\def\subsectionmark##1{}}

\def\@begintheorem#1#2{\it \trivlist \item[\hskip
 \labelsep{\bf #1\ #2.}]}
\def\@opargbegintheorem#1#2#3{\it \trivlist\item[\hskip%
 \labelsep{\bf #1\ #2.\ (#3)}]}
\def\@endtheorem{\endtrivlist}

\def\@listI{\leftmargin\leftmargini \parsep 1pt plus 2.5pt
 minus 1pt\topsep 5pt plus 2pt minus 3pt%
 \itemsep 0pt plus 2.5pt minus 1pt}
\let\@listi\@listI
\@listi

\def\@sect#1#2#3#4#5#6[#7]#8{\ifnum #2>\c@secnumdepth%
 \def \@svsec {}\else \refstepcounter {#1}\edef \@svsec%
 {\csname the#1\endcsname. \hskip .1em }\fi \@tempskipa%
 #5\relax \ifdim \@tempskipa >\z@ \begingroup #6\relax%
 \@hangfrom {\hskip #3\relax \@svsec }{\interlinepenalty%
 \@M #8.\par }\endgroup \csname #1mark\endcsname {#7}%
 \addcontentsline {toc}{#1}{\ifnum #2>\c@secnumdepth%
 \else \protect \numberline {\csname the#1\endcsname. }%
 \fi #7}\else \def \@svsechd {#6\hskip #3\@svsec #8.%
 \csname #1mark\endcsname {#7}\addcontentsline {toc}{#1}%
 {\ifnum #2>\c@secnumdepth \else \protect \numberline%
 {\csname the#1\endcsname. }\fi #7}}\fi \@xsect {#5}}

\def\section{\@startsection {section}{1}{\z@ }%
 {-3.5ex plus -1ex minus -.2ex}{2.3ex plus .2ex}{\bf }}

\def\thebibliography#1{%
 \section *{References.\@mkboth {REFERENCES}{REFERENCES}}%
 \list {[\arabic {enumi}]}{\settowidth \labelwidth {[#1]}%
 \leftmargin \labelwidth \advance \leftmargin \labelsep %
 \usecounter {enumi}} \def \newblock %
 {\hskip .11em plus .33em minus -.07em} \sloppy \clubpenalty 4000%
 \widowpenalty 4000 \sfcode`\.=1000\relax}

\def\@maketitle{%
 \newpage \null \vskip 2em
 \begin{center}{\Large\bf \@title \par }
 \vskip 1.5em
 {\large \lineskip .5em
 \begin {tabular}[t]{c}\@author
 \end{tabular}\par } \vskip .8em {June 20, 1994}
 \end{center}\par \vskip 1.5em}

\def\footnote{\@ifnextchar [{\@xfootnote }{\stepcounter {\@mpfn }%
\begingroup \let \protect
\noexpand \xdef \@thefnmark {\hskip-3mm}%
\endgroup \@footnotetext
}}
\def\@makefnmark{}

\def\abstract{%
\if@twocolumn \section *{Abstract}
 \else \small\quotation\noindent{\bf Abstract.}\fi}

\catcode`\@=13

\def\qed{\hspace*{\fill}
\mbox{\hphantom{mm}\rule{0.25cm}{0.25cm}}\\}

\begin{document}
\pagestyle{myheadings}
\bibliographystyle{plain}
\baselineskip20pt plus 2pt minus 1pt
\parskip3pt plus 1pt minus .5pt

\def\tilde#1{\widetilde{#1}}
\def\vect{\vec}
\def\oN{{\buildrel\circ\over N}}
\def\oU{{\buildrel\circ\over U}}
\def\dot{\buildrel \circ \over}

\def\P{{\cal P}} \def\ov{{\overline v}} \def\m{{\bf m}}
\def\V{{\cal V}} \def\oh{{\overline h}} \def\N{{\cal N}}
\def\T{{\cal T}} \def\ow{{\overline w}} \def\e{{\bf e}}
\def\W{{\cal W}} \def\F{{\cal F}} \def\D{{\cal D}}
\def\H{{\cal H}} \def\M{{\cal M}} \def\O{{\bf O}}
\def\Q{{\cal Q}} \def\B{{\cal B}}
\def\varomega{{\varphi}} \def\susp{{\uparrow}}
\def\bfR{{\bf R}} \def\bfRgeq{{\bf R}_{\geq 0}}
\def\bfRge{{\bf R}_{> 0}}
\def\sfF{{\sf F}}
\def\usfF{{\underline {\sf F}}}
\def\osfF{{\buildrel \circ \over {{\sf F}}}}
\def\oY{{\buildrel \circ \over Y}}
\def\oZ{{\buildrel \circ \over Z}}
\def\oW{{\buildrel \circ \over W}}
\def\u#1{{U[#1]}}\def\w#1{{W[#1]}}

\def\Conf#1#2{{C^0_{#1}({#2})}} \def\Confrm#1{{C^0_{{#1}}({\bf R}^m)}}
\def\FConf#1#2{{FC^0_{#1}({#2})}}
\def\CompConf#1#2{{C_{#1}({#2})}}
\def\CompFConf#1#2{{FC_{#1}({#2})}}
\def\uCompFConf#1#2{{{\underline FC}_{#1}({#2})}}
\def\conf#1{{C^0_{#1}({\bf R}^m)}}
\def\confn #1{{NC^0_{#1}({\bf R}^m)}}
\def\coll#1#2{{\{#1(n)\}_{n\geq #2}}}
\def\Aff{{\mbox{\rm Aff}}}
\def\Tbin{{\cal T}^{\geq 2}}
\def\MTbin{{M\cal T}^{\geq 2}}
\def\Tbine{{\cal T}^{\geq 2,e}}
\def\MTbine{{M\cal T}^{\geq 2,e}}
\def\inp{{{\rm inp}}} \def\edg{{{\rm edg}}}
\def\vert{{{\rm vert}}}

\begin{center}
{\Large \bf
A compactification of the real configuration space}
\end{center}
\begin{center}
{\Large \bf as an operadic completion}
\end{center}

\begin{center}
{\large Martin Markl}
\end{center}

\footnote{This work was supported by a Fulbright grant.}

\section{Introduction and summary}
\label{sec1}

For a compact Riemannian manifold $V$, Axelrod and Singer constructed
in~\cite{AS} a compactification $C_n(V)$ of the configuration space
$C^0_n(V)$ of $n$ distinct points in $V$, by adding to $C^0_n(V)$
the blowups along the
diagonals. Their construction works also for a noncompact manifold
$V$. In this case
the resulting object will not be compact (the configurations that
approach
spatial infinity have no limit), so it would perhaps be
better to speak about the `resolution of diagonals' rather than about
a
`compactification', as was done in~\cite{Gin}, but we will respect the
vicissitudes of history and call the process a `compactification'.

There is another, similar compactification of the {\em moduli space\/}
$\osfF_m(n)$
of configurations of $n$ distinct points in the $m$-dimensional
Euclidean
plane ${\bf R}^m$ modulo the action of the affine group,
described by Getzler and
Jones in~\cite{GJ} and denoted by $\sfF_m(n)$. The authors
of~\cite{GJ} also
stated that the collection $\sfF_m := \{\sfF_m(n)\}_{n\geq 1}$ has a
natural structure of a topological operad. This was a well known
fact for $m=1$,
because the collection $\sfF_1 = \{\sfF_1(n)\}_{n\geq 1}$
is nothing else but the
operad $K = \{K_n\}_{n\geq 1}$
of the `associahedra' introduced by J.~Stasheff in
his work~\cite{St} on homotopy associative spaces. Let us remark that
for
$m=2$ the
operad $\sfF_m$ plays an important r\^ole in topological closed string
field
theory.

The compactification $C_n(S^1)$ of the configuration space
$C^0_n(S^1)$ of $n$ distinct points on
the circle was studied by Bott and Taubes in~\cite{BT} as the basic
tool for
the construction of `nonperturbative' link invariants. It obviously
admits a
free $S^1$-action and the quotient $W_n := C_n(S^1)/S^1$ is what
J.~Stasheff
called in~\cite{St1} the `cyclohedron'. In the same paper he observed
that
the collection $W := \{ W_n \}_{n\geq 1}$ has a natural
structure of a right module over
the operad $K = \sfF_1$ in the sense introduced by us
in~\cite[page~1476]{models}.

The first aim of this work is to generalize this statement to the case
of an
arbitrary $n$-dimensional
Riemannian manifold $V$, i.e.~to prove that the collection $C(V) =
\{C_n(V)\}_{n\geq 1}$
has a natural structure of a right module over the operad of the
compactification of the moduli space of `local configurations'
$\sfF_n$.
Strictly speaking, this is true only for parallelizable manifolds, but
even
this class contains nontrivial and relevant examples, as we will see
later.
In the general case we must work with the {\em framed\/} version of
the
compactification, which we introduce
in~(\ref{Andulka-pusinka}).
The existence of the above mentioned structures has far-reaching
implications
to the geometry and combinatorics of the underlying spaces. We will
discuss
these questions in a forthcoming paper(s), see also the work of
M.~Ginzburg
and A.A.~Voronov~\cite{GV}.

We present an entirely new, purely algebraic construction of the
compactification based on the fact that
configuration spaces have a natural structure
of a partial operad (or a partial module over a partial operad, but we
will
not spoil the picture now).
We show that each partial operad admits an `operadic completion' and,
by a
miracle, this completion shows up to be the compactification we are
looking
for!

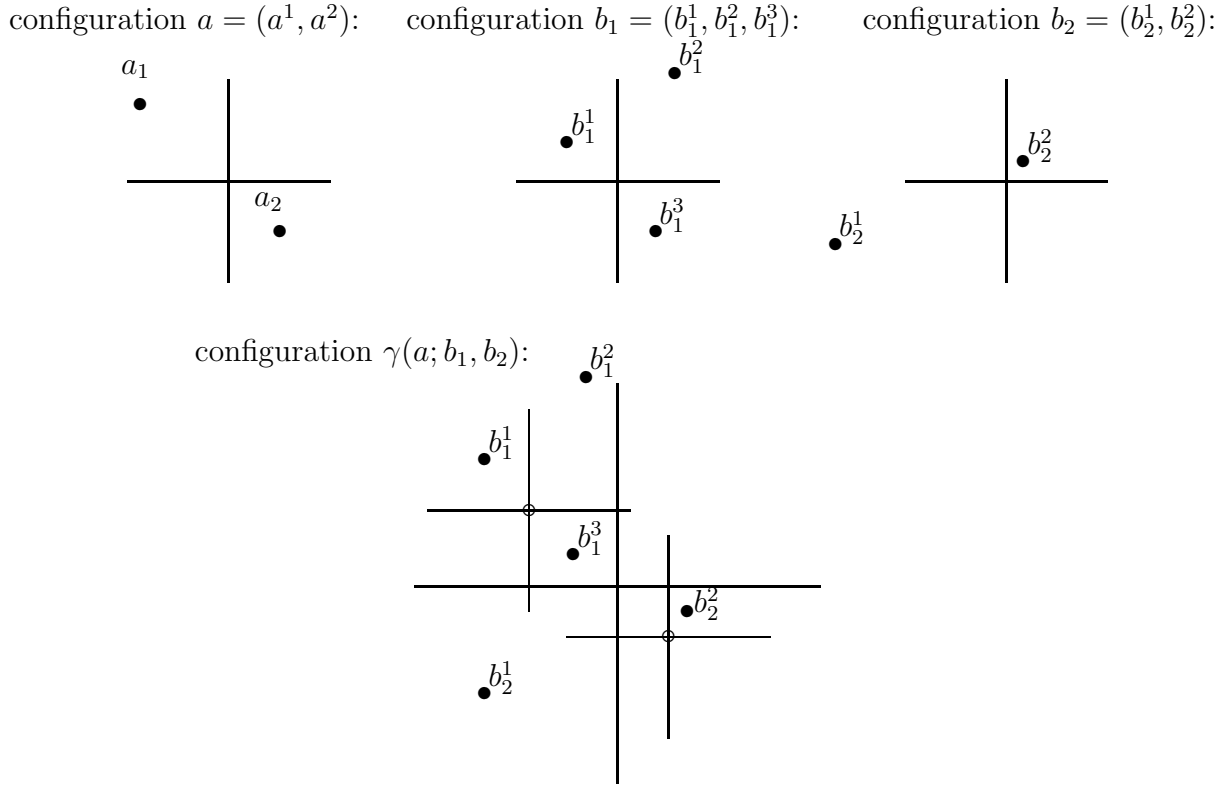
\begin{figure}
\label{Couperin}
\begin{center}
\setlength{\unitlength}{0.00045in}%

\begin{picture}(9687,8045)(1326,-8098)
\thicklines
\put(-1000,0){\put(1201,-1561){\line(1, 0){2400}}
 \put(3601,-1561){\line(-1, 0){1200}}
 \put(2401,-1561){\line( 0, 1){1200}}
 \put(2401,-361){\line( 0,-1){2400}}
 \put(1126,-286){\makebox(0,0)[lb]{$a_1$}}
 \put(2701,-1861){\makebox(0,0)[lb]{$a_2$}}
 \put(1351,-661){\makebox(0,0){$\bullet$}}
 \put(3001,-2161){\makebox(0,0){$\bullet$}}}
\put(4801,-1561){\line( 1, 0){2400}}
\put(7201,-1561){\line(-1, 0){1200}}
\put(6001,-1561){\line( 0, 1){1200}}
\put(6001,-361){\line( 0,-1){2400}}
\put(1000,0){\put(8401,-1561){\line( 1, 0){2400}}
 \put(10801,-1561){\line(-1, 0){1200}}
 \put(9601,-1561){\line( 0, 1){1200}}
 \put(9601,-361){\line( 0,-1){2400}}
 \put(7576,-2311){\makebox(0,0){$\bullet$}}
 \put(7776,-2111){\makebox(0,0){$b^1_{2}$}}
 \put(9800,-1336){\makebox(0,0){$\bullet$}}
 \put(10000,-1136){\makebox(0,0){$b^2_{2}$}}}
\put(3601,-6361){\line( 1, 0){4800}}
\put(6001,-3961){\line( 0,-1){4725}}
\thinlines
\put(5401,-6961){\line( 1, 0){2400}}
\put(7801,-6961){\line(-1, 0){1200}}
\put(6601,-6961){\line( 0, 1){1200}}
\put(6601,-5761){\line( 0,-1){2400}}
\put(3751,-5461){\line( 1, 0){2400}}
\put(6151,-5461){\line(-1, 0){1200}}
\put(4951,-5461){\line( 0, 1){1200}}
\put(4951,-4261){\line( 0,-1){2400}}
\thicklines

\put(4951,-5461){\makebox(0,0){$\circ$}}
\put(6601,-6961){\makebox(0,0){$\circ$}}

\put(6826,-6661){\makebox(0,0){$\bullet$}}
\put(7056,-6561){\makebox(0,0){$b^2_{2}$}}

\put(5401,-1111){\makebox(0,0){$\bullet$}}
\put(5601,-911){\makebox(0,0){$b^1_{1}$}}

\put(6676,-286){\makebox(0,0){$\bullet$}}
\put(6876,-86){\makebox(0,0){$b^2_{1}$}}

\put(6451,-2161){\makebox(0,0){$\bullet$}}
\put(6651,-1961){\makebox(0,0){$b^3_{1}$}}

\put(5626,-3886){\makebox(0,0){$\bullet$}}
\put(5826,-3686){\makebox(0,0){$b^2_{1}$}}

\put(4426,-4861){\makebox(0,0){$\bullet$}}
\put(4626,-4661){\makebox(0,0){$b^1_{1}$}}

\put(5476,-5986){\makebox(0,0){$\bullet$}}
\put(5676,-5786){\makebox(0,0){$b^3_{1}$}}

\put(4426,-7636){\makebox(0,0){$\bullet$}}
\put(4626,-7436){\makebox(0,0){$b^1_{2}$}}

\put(-1200,239){\makebox(0,0)[lb]{configuration $a = (a^1,a^2)$:}}
\put(3501,239){\makebox(0,0)[lb]{configuration $b_1 =
(b_1^1,b_1^2,b_1^3)$:}}
\put(8901,239){\makebox(0,0)[lb]{configuration $b_2 =
(b_2^1,b_2^2)$:}}
\put(1000,-3686){\makebox(0,0)[lb]{configuration
$\gamma(a;b_1,b_2)$:}}
\end{picture}
\end{center}
\caption{The partial operad structure on the configuration space made
easy.
The construction of $\gamma(a;b_1,b_2) \in \Conf 5{{\bfR}^2}$ from $a
\in
\Conf 2{{\bfR}^2}$, $b_1\in \Conf 3{{\bfR}^2}$ and
$b_2 \in \Conf 2{{\bfR}^2}$.}
\end{figure}

Let us try to give the reader a flavour how this partial operad
structure
looks. Consider the space $\Confrm n$ of configurations of $n$
distinct points in the Euclidean plane ${\bfR^m}$. To define an operad
structure on the collection $\Confrm{} = \{\Confrm{n}\}_{n\geq 1}$ we
need to
specify, for each $a = (a^1,\ldots,a^l) \in \Confrm l$ and $b_i
\in \Confrm {m_i}$, the value of the `composition map'
$\gamma(a;b_1,\ldots,b_l) \in \Confrm{m_1+\cdots+m_l}$. This can be
done by
putting
\[
\gamma(a;b_1,\ldots,b_l) :=
((\underbrace{a^1,\ldots,a^1}_{m_1{\ \rm times}})+b_1,
(\underbrace{a^2,\ldots,a^2}_{m_2{\ \rm times}})+b_2,\ldots,
(\underbrace{a_l,\ldots,a_l}_{m_l{\ \rm times}})+b_l).
\]
The
configuration $\gamma(a;b_1,\ldots,b_l)$ may be viewed as
the superposition of the configurations
$T_{a_1}(b_1),\ldots,T_{a_l}(b_l)$,
where $T_a(-)$ means, just here and now,
the translation by a vector $a \in {\bfR}^m$. This process is
visualized on
Figure~\ref{Couperin}.

We encourage the reader to verify that all the axioms of an operad are
satisfied. The only small drawback is that $\gamma(a;b_1,\ldots,b_l)$
need
not necessarily be an element of the configuration space
$\Confrm{m_1+\cdots+m_l}$, because the components of
$\gamma(a;b_1,\ldots,b_l)$ need not be different. Thus the structure
map
is defined only for {\em some\/} elements of $\Confrm l \times
\Confrm{m_1}
\times \cdots \times \Confrm{m_l}$; we will call such an object
a {\em partial operad\/}, though the definition we use is more subtle
and
differs a bit from the standard definition of a partial operad.

As far as we know, nobody has observed the existence of this partial
operad
structure before. It is implicitly hidden in the formulas
of~\cite[pages~25--29]{AS},
and, in fact, all this paper is based on a very
meticulous study of these pages.

As above, we will respect the notations introduced in~\cite{GJ}
resp.~\cite{BT,Th} despite their obvious incompatibility, i.e.~we will
use
the notation $\osfF_m(n)$ for the moduli space of configurations of
$n$
distinct points in ${\bfR}^m$ modulo the affine group action (=
dilatations
and translations), and $\Conf nV$ for the space of configurations
of $n$ distinct points in a manifold $V$.

\noindent
{\bf Summary of the paper.}
In the following section we explain our concept of a partial operad
and
construct an operadic completion of such an object. In
Section~\ref{sec3} we
define a partial operad of virtual configurations $\chi$ and a framed
version $f\chi$ of this object. We show that the operadic completion
$\tilde
\chi$ (resp.~$\widetilde {f\chi}$) of $\chi$ (resp.~$f\chi$) coincides
with
the compactification $\sfF_m$ (resp.~the
framed version $f\sfF_m$) considered by Getzler and Jones
in~\cite{GJ}. This
immediately implies the existence of an operad structure on these
objects.

In Section~\ref{sec4} we introduce our notion of partial modules over
a
partial operad and describe a module completion of these objects. In
Section~\ref{sec5} we define,
for each Riemannian manilfold $V$, the partial module
of framed virtual configurations $f\mu$ (resp.,~if $V$ is
parallelizable, the
partial module of virtual configurations $\mu$). We show that the
module
completion $\widetilde {f\mu}$ (resp.~$\tilde \mu$) coincides with the
Axelrod-Singer compactification $\CompFConf {}V$
(resp.~$\CompConf{}V$). As an
immediate consequence we see that $\CompFConf {}V$ is a natural right
module
over the operad $f\sfF_m$ (resp.~that $\CompConf {}V$ is a natural
right module
over $\sfF_m$). Observe that there is many parallelizable manifolds
for which
the configuration spaces are interesting, for example the spheres $V
=S^m$,
for $m = 1,3,7$. Another important case is $V = {\bfR}^m$ or $V = $
the torus,
or, still more generally, $V =$ a (not necessary compact) Lie group.

In the last section we exploit the well-known
fact that the above mentioned
compactifications are manifold with corners. We get immediately, from
a
result of J.~Cerf~\cite{Cerf}, that each of those compactifications is
diffeomorphic to a closed submanifold obtained by a truncation of its
open
part. An explicit construction of such a truncation was given, for $K
=
\sfF_1$,
by S.~Sternberg and S.~Shnider
in~\cite{SS}; the authors show that the associahedron can be
constructed as a truncation of the $(n-1)$ dimensional simplex
$\Delta^{n-1}$. The possibility of a similar construction of the
cyclohedron
was observed in the appendix to~\cite{St1} by J.~Stasheff.

For any manifold with corners $M$, the skeletal filtration induces a
spectral
sequence. As suggested by~\cite[Lemma~3.4]{GJ}, the first term of this
spectral
sequence can be, for $M=$ one of the compactifications above,
identified to the bar construction (or a suitable
generalization) over an operad (or a module) formed by the cohomology
of the `open parts' of these spaces. Our
approach gives a straightforward definition of these operad
structures,
more direct that the standard one based on a chain of homotopy
equivalences
with a little-disks-type object. This gives us a very easy
understanding of the first term of this spectral sequence.

\noindent
{\bf Acknowledgements.}
I would express my thanks to Jim Stasheff for numerous discussions and
encouragement. Also the communication with Sasha Voronov, who was
working
independently on~\cite{GV}, was very useful.

\section{Algebraic background I}
\label{rucicka}

\begin{odstavec}\label{trees}
Language of trees.{\rm\
Let $\T_n$ denote the set of all (rooted, connected)
trees with $n$ input edges. For
such a tree
$T\in \T_n$, let $\vert(T)$ denote the set of its vertices.
For $v\in \vert(T)$,
let $\inp(v)$ be the set
of input edges of $v$; $\inp(v)$ will sometimes denote also the number
of input edges of $v$, the meaning will always be clear from the
context.
The set $\T_n$ of all $n$-trees has a natural partial order; we say
that
$S\leq T$ if the tree $S$ was obtained from $T$ by collapsing one or
more of
its inner edges. The set $\T_n$ has a unique minimal element $T(n)$,
the
$n$-corolla, the tree with exactly one vertex.
}\end{odstavec}

\begin{odstavec}\label{collections}
Collections and operads.{\rm\
Recall that a (topological) {\em collection\/} is just a sequence $E =
\coll
E1$ of topological spaces.
For a collection $E$ and a tree $T \in \T_n$,
let $E(T)$ denote the set of all colorings of
the vertices of $T$ by elements of $E$ such that a vertex $v$
is colored by an element from $E(\inp(v))$; observe that $E(T(n)) =
E(n)$.
For $v\in \vert(T)$ and $\xi \in
E(T)$, denote by $\xi(v)$ the value of the coloring $\xi$ at $v$.
If $T,S \in \T_n$, $S\leq T$,
then each $h\in \H := \vert(S)$ labels a
subtree $T_h$ of $T$ whose vertices collapsed to
$h$.

For each collection $E$ there exists the
{\em free operad\/} $\F(E)$ generated by $E$. As a collection, it is
defined by
\[
\F(E)(n) = \coprod_{T\in \T_n} E(T),
\]
while the operad structure is given by the grafting of the underlying
trees.
Let us recall~\cite{GK} that an operad structure on
a collection $E$ can be defined by specifying, for each $T \in \T_n$,
a
structure map
$\gamma_T : E(T) \to E(T(n))=E(n)$;
these maps must behave well under the grafting
operation of underlying trees.
More precisely, let $S\leq T \in \T_n$. Then the restriction defines,
for
each $h \in \H := \vert(S)$, the map $r_h :E(T) \to E(T_h)$.
Let us introduce the `operadic extension' $\{\gamma_{S,T}\}_{S\leq T}$
of the
system $\{\gamma_T\}_T$, $\gamma_{S,T}:E(T) \to E(S)$, by
\begin{equation}
\label{operadic-extension}
\gamma_{S,T}(\xi)(h) := \gamma_{T_h}(r_h(\xi)),\ h \in \vert(S),
\end{equation}
see~\ref{trees} for the notation. Then we
require that
\begin{equation}
\label{axiom}
\gamma_T(\xi) = \gamma_S(\gamma_{S,T}(\xi)),
\mbox{ for each $\xi\in E(T)$, $S,T \in \T_n$, $S\leq T$}.
\end{equation}
}\end{odstavec}

\begin{odstavec}\label{partial-operads}
Partial operads.{\rm\
We say that a
{\em partial operad\/} is a collection $E$
with structure maps defined only on
a subset $\u T$ of $E(T)$, $\gamma_T : \u T \to E(n)$, $T \in \T_n$.
These maps
are supposed to satisfy~(\ref{axiom}) whenever the
corresponding compositions are defined. To understand this better, we
introduce the set
\begin{equation}
\label{Mikinka}
U_S(T) := \{ \xi \in E(T);\
r_h(\xi) \in \u{T_h},\ h\in \vert(S)
\} \subset E(T).
\end{equation}
Observe that $U_S(S) = U(S)$, the set of all colorings of the tree $S$
by
elements of the collection $U := \coll U1$ with $U(n) = \u{T(n)}$,
while the
opposite extreme is $U_{T(n)}(T) = \u T$. The sets $U_S(S)$
will play the r\^ole of `open strata' and we denote them by $R_S$.

The map $\gamma_{S,T}$ on the right side of~(\ref{axiom}) is defined
for
$\xi \in U_S(T)$, we thus require~(\ref{axiom}) to be satisfied only
for
\begin{equation}
\label{paska}
\xi \in \u T \cap U_S(T) \mbox{ such that } \gamma_{S,T} (\xi) \in \u
S.
\end{equation}

Let $\P = (E = \coll E1, \{\gamma_T :\u T \to E(n)\})$ be a
partial operad. We make life easier by assuming that
\begin{equation}
\label{21}
\mbox{\rm Im}(\gamma_T) \subset \u{T(n)}=: U(n),\ T \in \T_n,
\end{equation}
as our basic examples will always share this property.
Let
\begin{equation}
\label{Flicek}
\hat U(T) := \bigcup_{S\leq T} U_S(T),
\end{equation}
{\em topologized as a subset of $E(T)$\/}. Consider the collection
$\hat U =
\coll {\hat U}1$ defined by
\[
\hat U(n) := \coprod_{T\in \T_n} \hat U(T)\mbox{ (disjoint union)}.
\]
}\end{odstavec}
\begin{lemma}
\label{Pik}
The collection $\hat U =
\coll {\hat U}1$ is a topological suboperad of the free operad $\F =
\F(E)$.
\end{lemma}

\noindent
{\bf Proof.}
The proof is almost immediate.
Let $\xi \in U_S(T)$, $S \leq T \in T_l$, and $\xi_i \in
U_{S_i}(T_i)$, $S_i
\leq T_i \in \T_{m_i}$, $1\leq i\leq l$. Let
$\gamma(T;T_1,\ldots,T_l)$
(resp.~$\gamma(S;S_1,\ldots,S_l)$) denote the tree obtained by
grafting the
tree $T_i$ at the $i$-th input of $T$ (resp.~the
tree $S_i$ at the $i$-th input of $S$), for $1\leq i \leq l$.
Clearly $\gamma(S;S_1,\ldots,S_l) \leq
\gamma(T;T_1,\ldots,T_l) \in \T_{m_1+\cdots+
m_l}$.
If $\gamma_{\F}$ denotes the composition map of the free operad $\F$,
we immediately see that
\[
\gamma_{\F}(\xi;\xi_1,\ldots,\xi_l) \in
U_{\gamma(S;S_1,\ldots,S_l)}(\gamma(T;T_1,\ldots,T_l))
\]
which finishes the proof.\qed

Condition~(\ref{21}) implies that the map $\gamma_{S,T}$ introduced
in~(\ref{operadic-extension}) maps $U_S(T)$ to $U_S(S) = R_S$.
We may thus define $\tilde P = \coll {\tilde P}1$ by
\begin{equation}
\label{Alicek}\label{Kacirek}
\tilde P (n) := \hat U(n)/\sim
\end{equation}
where the relation $\sim$ identifies elements $\xi$ of $U_S(T)$ with
their
images $\gamma_{S,T}(\xi) \in R_S \subset E(S)$. In the following
proposition, $U = \coll U1$ is the collection defined in~(\ref{21})
and
$\F(U)$ is the free operad generated by this collection.

\begin{proposition}
\label{casio}
The collection $\tilde \P = \coll {\tilde P}1$ is a topological
operad.
There exists a natural epimorphism of topological operads
\[
\rho: \F(U) \to \tilde \P.
\]
If the sets $U_S(T)$ are `combinatorially independent' in the sense
that
\begin{equation}
\label{kralicek}
S',S'' \leq T,\ S'\not= S'' \Longrightarrow U_{S'}(T) \cup U_{S''}(T)
=
\emptyset,
\end{equation}
then the map $\rho$ is an isomorphism of sets.
\end{proposition}

In the light of the proposition,
we may view $\tilde \P(n)$ as obtained by glueing the `open strata'
$U(T) = R_T$, $T\in \T_n$, of $\F(U)(n)$
in a way compatible with the operad structure.
We call $\tilde \P$ the
{\em operadic completion\/} of the partial operad $\P$.

\noindent
{\bf Proof of the proposition.}
We prove that the operad structure on $\hat U$ induces an operad
structure on its quotient~(\ref{Alicek}). To this end,
we must show that the equivalence
$\sim$ is compatible with the operad structure on $\hat U$. This is,
however,
evident; we defined the system $\{\gamma_{S,T}\}_{S\leq T \in \T_n}$
by extending
$\{\gamma_T\}_{T\in \T_n}$ as operad maps, and the claim follows from
the
definition of $\sim$.

As for the second part of the theorem, the inclusion $\iota : U \to
\hat U$
of collections given by
\[
\iota(n) : U(n) = \u{T(n)} = U_{T(n)}(T(n)) \hookrightarrow \hat U(n)
\]
extends to a continuous map $\rho :\F(U) \to \tilde \P$, by the
freeness of
the operad $\F(E)$.
The very definition of the relation $\sim$ implies that each $\xi \in
\hat
U(n)$ is equivalent to some $\xi' \in R_S \subset \mbox{Im}(\rho)$.
This
implies that the map $\rho$ is an epimorphism. The independence
condition~(\ref{kralicek})
then assures that the relation $\sim$ cannot identify
two distinct points of $R_S$, which shows that $\rho$ is a
monomorphism.\qed

\section{Compactification of the moduli space}
\label{sec3}

\begin{odstavec}
\label{virtual-configurations}
{\rm\
We open this section by defining the
{\em partial operad of virtual configurations\/}
$\chi = (E,\{\gamma_T :\u{T}
\to E(n)\})$. The collection $E$
is given by
$E(n) := [\bfRgeq \times \osfF_m(n)]$ for $n\geq 2$,
while $E(1) = \emptyset$. We must also specify, for each
$T\in \T_n$, a subset $\u{T} \subset E(T)$ and a composition map
$\gamma_T : \u{T} \to E(n)$. Since $E(1) = \emptyset$, the set $E(T)$
can
be nonempty only for trees all of whose vertices have
at least two input edges; we denote the set of all such trees by
$\Tbin_n$.

First of all, an element of $E(T)$ is a sequence
\begin{equation}
\label{10}\label{notation}
\xi = \{\kappa_w;\ w\in \W\},\
\kappa_w = (t_w,[\vect z_w])
\in[ {\bf R}_{\geq 0} \times \osfF_m(\inp(w))] \mbox{ for }w\in \W =
\vert(T).
\end{equation}
We can assume that the vectors
$\vect z_w =(z^1_w,\ldots,z_w^{i_w})$, where $i_w := \inp(w)$,
are {\em normalized\/}
in the sense that
\begin{equation}
\label{norm}
\sum_{1\leq i\leq i_w} z^i_w = 0 \mbox{ and }
\sum_{1\leq i\leq i_w} |z^i_w|^2 = 1,
\end{equation}
where $|\!-\!|$ denotes the Euclidean norm in $\bfR^m$.

Let $\ow$ be the terminal vertex of the tree $T$ and
let $Y(T) \subset E(T)$ be the set of all elements as in~(\ref{10})
such
that $t_{\ow} = 0$.
As the first step towards $\gamma_T$ we
define, for $T \in \Tbin_n$,
a map $\omega_T : Y(T) \to (\bfR^m)^n$ as follows.

For any $1\leq i\leq n$ there exists in $T$ a unique
path from the $i$-th input to the output, as in Figure~\ref{picture2}.
\begin{figure}
\begin{center}
\setlength{\unitlength}{0.00095in}%

\begin{picture}(6324,1149)(1189,-3448)
\thicklines
\put(1201,-2761){\vector( 1, 0){1200}}
\put(2401,-2761){\vector( 1, 0){1200}}
\put(3601,-2761){\vector( 1, 0){1200}}
\put(6301,-2761){\vector( 1, 0){1200}}
\put(2026,-2386){\vector( 1,-1){375}}
\put(1951,-2536){\vector( 2,-1){450}}
\put(1951,-3361){\vector( 3, 4){450}}
\put(3151,-2311){\vector( 1,-1){450}}
\put(3001,-2461){\vector( 2,-1){600}}
\put(3151,-3436){\vector( 2, 3){450}}
\put(4501,-2311){\vector( 2,-3){300}}
\put(4201,-2311){\vector( 4,-3){600}}
\put(4351,-3436){\vector( 2, 3){450}}
\put(5101,-2761){\vector( 1, 0){1200}}
\put(5701,-2311){\vector( 4,-3){600}}
\put(5401,-2311){\vector( 2,-1){900}}
\put(5851,-3436){\vector( 2, 3){450}}
\put(2401,-2761){\makebox(0,0){$\bullet$}}
\put(3601,-2761){\makebox(0,0){$\bullet$}}
\put(4801,-2761){\makebox(0,0){$\bullet$}}
\put(4930,-2760){\makebox(0,0){$\cdots$}}
\put(6301,-2761){\makebox(0,0){$\bullet$}}
\put(6300,-3000){\makebox(0,0)[lb]{$w_k = \ow$}}
\put(3600,-3000){\makebox(0,0)[lb]{$w_2$}}
\put(2400,-3000){\makebox(0,0)[lb]{$w_1$}}
\put(1490,-2650){\makebox(0,0)[b]{$r_1$-th input}}
\put(2840,-2650){\makebox(0,0)[b]{$r_2$-th input}}
\put(4090,-2650){\makebox(0,0)[b]{$r_3$-th input}}
\put(5440,-2650){\makebox(0,0)[b]{$r_k$-th input}}
\end{picture}
\end{center}
\caption{
\label{picture2}
A path in the tree $T$.
}
\end{figure}
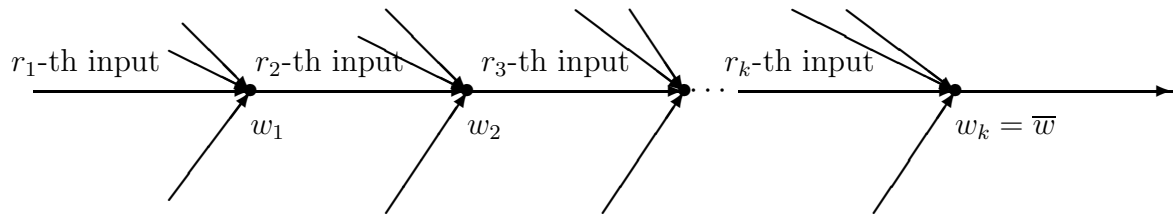
Using the notation above, we put
\begin{equation}
\label{fr}
\omega_i(\xi):= z_{\ow}^{r_k} + t_{w_{k-1}} \cdot
z_{w_{k-1}}^{r_{k-1}} +
\cdots + t_{w_1}\cdots t_{w_{k-1}} \cdot z^{r_1}_{w_1}
\end{equation}
and, finally, $\omega_T(\xi) := (\omega_1(\xi),\ldots,\omega_n(\xi))$.
The following observation is interesting and we formulate it
though we will not need it in the sequel; the proof is immediate.
}\end{odstavec}

\begin{observation}
The map $\omega_T : Y(T) \to (\bfR_{\geq 0}^m)^n$ is a monomorphism,
for any
$T \in \Tbin_n$.
\end{observation}

Let $\u{T}$ be the set of all $\xi \in Y(T)$ such that all the points
$\omega_1(\xi),\ldots, \omega_n(\xi) \in \bfR^m$ are distinct.
Then $\gamma_T: \u{T} \to E(n)$ is defined as
the composition of the restriction $\omega_T|_{\u{T}}$ with the
projection $\Conf n{{\bfR}^m} \to \osfF_m(n)$ and the inclusion
$\osfF_m(n) =
\{0\} \times\osfF_m(n) \hookrightarrow E(n)$.

\begin{proposition}
\label{object}
The object
\[
\chi = (E=\{E(n)\}_{n \geq 1},\{\gamma_T :\u{T} \to
E(n)\}_{T\in \Tbin_n}\})
\]
defined above is a partial operad satisfying the independence
condition~(\ref{kralicek}).
\end{proposition}

\noindent
{\bf Proof.}
Let us prove the combinatorial independence first.
If $\xi \in U_S(T)$ is as
in~(\ref{notation}) then, by definition, $r_h(\xi) \in \u{T_h}$ for
all
$h\in \H$, see~(\ref{Mikinka}).
This clearly implies that $t_w = 0$ if and only if
$w$ is the output vertex of some $T_h$. Thus the set $\{ w\in
\vert(T);\ t_w
= 0\}$ uniquely determines a tree $S$ with $S\leq T$ such that $\xi
\in
U_S(T)$; the combinatorial independence is now obvious.

We must of course verify also that $\chi$ is a partial operad.
But this is easy: the independence implies that, if
$U_S(T) \cap \u{T} \not= \emptyset$, then
$S = T(n)$, the $n$-corolla. Thus, by~(\ref{paska}),
the only thing which has
to be verified is the unitarity,
$\gamma_{T(n)} = \mbox{id}$, which is immediate from
the definition.\qed

\begin{theorem}
\label{mys}
The operad completion
$\tilde \chi$ of the partial operad $\chi$
coincides with
the compactification $\sfF_m$ of the moduli space of points in the
plane discussed in~{\rm\cite{GJ}}, $\sfF_m(n) =
\tilde \chi(n)$ for any $n\geq 1$.
\end{theorem}

\noindent
{\bf Proof.}
We prove the theorem by constructing an explicit isomorphism $z:
\tilde
\chi(n) \to \sfF_m(n)$.
Let $\xi = \{ \kappa_w = (t_w,[\vect z_w]);\ w\in \W\} \in U_S(T)
\subset
E(T)$ be a point as in~(\ref{notation}).
As we already saw in the proof of
Proposition~\ref{object},
the tree $S$ uniquely determines a subset $\W_S \subset \W$;
$\W_S := \{ w\in \W;\ t_w = 0\}$. For each $\epsilon > 0$ define
$\tau_\xi(\epsilon) \in E(T)$ by
\[
\tau_\xi(\epsilon)= \{
\lambda_w = (s_w,[\vect z_w]);\ w\in \W\},
\]
where $s_w := t_w$ for $w\in \W \setminus \W_S$, and $s_w := \epsilon$
for $w
\in \W_S$. By \cite[lemma in \S5.4]{AS}, $\tau_\xi(\epsilon)
\in \u T$ for small $\epsilon$. Thus, for small $\epsilon$,
$\gamma_T(\tau_\xi(\epsilon))$ is a curve in $\osfF_m(n)$ which
converges,
for $\epsilon \to 0$, to a point in the compactification $\sfF_m(n)$.
We
define
\[
z(\xi):=
\lim_{\epsilon \to 0}(\gamma_T(\tau_\xi(\epsilon)))
\mbox{ in } \sfF_m(n).
\]
We must prove that this definition is compatible with the defining
relation $\sim$
of~(\ref{Alicek}). This was in fact done in the proof of a theorem
in~\cite[\S5.4]{AS}, and our claim becomes clear if we compare our
$\gamma_{S,T}$ with the formulas~\cite[(5.77.1), (5.82)]{AS}
for the extension of
the map $\psi_0$, though the verification is rather difficult because
of the
difference between notations used. Fortunately, the claim can be
verified
more or less directly, if we realize what we are suppose to verify.

We have to verify the following. If $\xi$ is as above, let $\xi' :=
\gamma _{S,T}(\xi) \in U_S(S)$ and let $\tau_{\xi'}(\epsilon) \in
E(S)$ be
the corresponding curve. Then we must prove that
\[
\lim_{\epsilon \to 0}(\gamma_T(\tau_\xi(\epsilon))) =
\lim_{\epsilon \to 0}(\gamma_S(\tau_{\xi'}(\epsilon)))
\mbox{ in } \sfF_m(n).
\]
To do this, we must write explicit formulas for the curves
$\gamma_T(\tau_\xi(\epsilon)$ and $\gamma_S(\tau_{\xi'}(\epsilon)$ and
then
use a criterion of~\cite[\S5.2]{AS} to compare points in the
compactification
which are presented as limits of curves in the `open part'. This is a
straightforward, though not exactly easy, verification.\qed

\begin{odstavec}
{\rm\
We need also a `framed' version of the operad $\osfF_m$. It will be an
`$G$-operad' with $G = \O(n)$, where by and $G$-operad we mean an
operad
$\P = \coll{\P}1$ such that each $\P(n)$ is a (left)
$G$-space and the
composition map satisfies
\begin{equation}
\label{Vasik}
g(\gamma(x;x_1,\ldots,x_l)) = \gamma(gx;x_1,\ldots,x_l),\ x\in \P(l),\
x_i \in
\P(m_i),\ 1\leq i\leq l,\ g\in G.
\end{equation}
A typical example of such an object is the $\O(m)$-operad $f\D_m$ of
framed
little $m$-disks. More generally, suppose we have an (ordinary)
operad $\P$ such that each $\P(n)$ is a (left) $G$-space and such that
the
composition map satisfies, under the notation of~(\ref{Vasik}),
$g\gamma(x; x_1,\ldots,x_l) = \gamma(gx; gx_1,\ldots,gx_l)$. An
example
is the ordinary little $m$-disks operad $\D_m$ with the action of
$\O(m)$ induced by the representation of this group on the ambient
affine
space.
Then the operad $G\P$ with $G\P(n) := \P(n) \times G^{\times n}$, with
the
diagonal action of the group $G$ and the composition map $\gamma_G$
defined as
\begin{eqnarray*}
\lefteqn{
\gamma_G((x,g_1,\ldots,g_l); (x_1,g^1_1,\ldots,g^{m_1}_1),\ldots,(x_l,
g^1_l,\ldots,g^{m_l}_l)) :=\hskip2cm}
\\&&
\hskip2cm (\gamma(gx;
x_1,\ldots,x_l),g_1g^1_1,\ldots,g_1g^{m_1}_1,\ldots,
g_lg^1_l,\ldots,g_lg^{m_l}_l)
\end{eqnarray*}
is a $G$-operad in the sense of~(\ref{Vasik}).
We believe that the analogous notion of a {\em
partial\/} $G$-operad is clear.

We are going to define now, for any $m\geq 1$, the partial
$\O(m)$-operad
of framed virtual configurations $f\chi = (fE, \{ f\gamma_T : f\u T
\to
fE(n)\})$. Let $fE(n) := E(n) \times \O(n)^{\times n}$, where the
collection
$E(n)$ is the
same as in the definition of the partial operad $\chi$
in~\ref{virtual-configurations}.
A typical
element $\rho \in fE(T)$, $T \in \Tbin_n$, looks like
\begin{eqnarray*}
&\rho = \{\rho_w;\ w\in \W\},\ \rho_w =
(t_w,[\vect z_w],\vect g_w),\ w\in \W,\mbox{ with}&
\\
&t_w\in {\bfR}_{\geq 0},\ [\vect z_w] = [z_w^1,\ldots,z_w^{i_w}]
\in \osfF_m(i_w),\ \vect g_w = (g_w^1,\ldots,g_w^{i_w}) \in
\O(m)^{\times
i_w},\ i_w := \inp(w).&
\end{eqnarray*}
Let $fY = \{\rho \in fE(T);\ t_{\ow} = 0\}$, where $\ow$ is the output
vertex
op $T$. For
$1\leq i\leq n$, we define the framed version of the map $\omega_i$
of~(\ref{fr}) as
\begin{equation}
\label{veprik}
f\omega_i(\phi):= z_{\ow}^{r_k} + t_{w_{k-1}} \cdot
(g_{w_{k-1}}^{r_{k-1}}z_{w_{k-1}}^{r_{k-1}}) +
\cdots + t_{w_1}\cdots t_{w_{k-1}} \cdot
(g_{w_{k-1}}^{r_{k-1}} \cdots g^{r_1}_{w_1} z^{r_1}_{w_1}),
\end{equation}
where we use the same notation based on Figure~\ref{picture2}
as in~(\ref{fr}). As before, put
\[
f\u T = \{ \rho \in Y(T);\mbox{ the points $f\omega_1(\rho), \ldots,
f\omega_n(\rho)$ are distinct}\}.
\]
Finally, let $f\gamma_T(\rho) :=
[\omega(\rho)] \times (g_1,\ldots,g_n)$, where $[\omega(\rho)]$
denotes the
class of $\omega(\rho) \in \Conf n{{\bfR}^m}$ in $\osfF_m(n)$ and
$g_i := g_{w_{k}}^{r_{k}} \cdots g^{r_1}_{w_1}$, $1\leq i\leq n$.
We have the following `framed' version of Proposition~\ref{object}.
}\end{odstavec}

\begin{proposition}
The object
\[
f\chi = (fE=\{fE(n)\}_{n \geq 1},\{f\gamma_T :f\u T \to
fE(n)\}_{T\in \Tbin_n}\})
\]
defined above is a partial $\O(m)$-operad satisfying the independence
condition~(\ref{kralicek}). It contains $\chi$ as a natural suboperad.
\end{proposition}

It is well-known~\cite{GJ} that the compactification $\sfF_m(n)$ of
the
moduli space $\osfF_m(n)$ admits
a natural action of the group $\O(m)$. This action
can be used to introduce the
`framed' version $f\sfF_m(n)$ of
the space $\sfF_m(n)$ by $f\sfF_m(n) := \sfF_m(n)
\times \O(m)^{\times n}$, with the diagonal action of the group
$\O(n)$. We
have the following analog of Theorem~\ref{mys}.

\begin{theorem}
The operadic completion
$\widetilde{f \chi}$ of the partial $\O(m)$-operad $f\chi$
coincides with the framed version $f\sfF_m$ of the
the compactification $\sfF_m$ of the moduli space of points in the
plane introduced above, $f\sfF_m(n) =
\widetilde {f\chi}(n)$ for any $n\geq 1$.
\end{theorem}

\section{Algebraic background II}
\label{sec4}

\begin{odstavec}
Modules over operads.{\rm\
Let $M$ and $E$ be topological collections and $T\in \T_n$ a tree.
Denote by
$M_E(T)$ the set of all colorings of the tree $T$ such that the output
vertex
of $T$ is colored by an element of $M$ while the remaining vertices
are
colored by elements of $E$. Suppose that the collection $E$ forms an
operad
with the structure maps
$\{\gamma_T :E(T) \to E(n)\}$ as in~\ref{collections}.
One way to define on a collection $M$
a {\em right module structure\/} over
the operad $E$ in the sense of~\cite{models} is to specify maps
$\nu_T : M_E(T) \to M(n)$, $T\in \T_n$, which behave well
in the following sense, compare~\ref{collections}.

Let $T,S \in \T_n$, $S\leq T$. Let $\H := \vert(S)$ and let $\oh$ be
the
output vertex of the tree $S$. Decompose
$\H$ as $\H = \{\oh\} \cup \H'$.
The restriction gives the map $r_{\oh} :M_E(T) \to M(T_{\oh})$ and,
for each $g \in \H'$, the map $r_g : M_E(T)
\mapsto E(T_g)$. We define the `modular extension'
$\{\nu_{S,T}\}_{S\leq T}$
of the system $\{\nu_T\}_T$, $\nu_{S,T} :M_E(T) \to M_E(S)$, by
\begin{equation}
\label{modular-extension}
\nu_{S,T}(\eta)(\oh) := \nu_{T_{\oh}}( r_{\oh}(\eta))
\mbox{ and }
\nu_{S,T} (\eta)(g): = \gamma_{T_g}(r_g(\eta)),\
g\in \H'.
\end{equation}
Then we require that
\begin{equation}
\label{triska}
\nu_T(\eta) = \nu_S(\nu_{S,T}(\eta)),
\mbox{ for $\eta \in M_E(T)$, $S\leq T$, $S,T \in \T_n$}.
\end{equation}
}\end{odstavec}

\begin{odstavec}
Partial modules.{\rm\
Let $\P = (E,\gamma_T :\u T \to
E(n))$ be a partial operad as in~\ref{partial-operads}. Then a
structure of a {\em partial module\/} over a partial operad $\P$ will
be
given by specifying, for each $T\in \T_n$, a subset $\w T \subset
M_E(T)$ and
a map $\nu_T : \w T \to M(n)$ such that the maps $\{\nu_T\}_T$
satisfy~(\ref{triska}),
whenever the compositions involved are defined.
As in~\ref{partial-operads}
this means that we require~(\ref{triska}) only for $\eta \in
\w T \cap W_S(T)$ with $\nu_{S,T}(\eta) \in \w S$,
where $W_S(T)$ is the subset of $M_E(T)$ defined as
\begin{equation}
W_S(T) := \{ \eta \in M_E(T);\ r_{\oh}(\eta) \in \w{T_{\oh}} \mbox{
and }
r_g(\eta) \in \u{T_g},\ g\in \H'
\}.
\end{equation}
We suppose,
as in~(\ref{21}), that
\begin{equation}
\label{rest}
\mbox{\rm Im}(\nu_T) \subset \w{T(n)}=: W(n),\ T \in \T_n.
\end{equation}

Let $\M = (M = \coll M1,\{\nu_T :\w T \to M(n)\})$ be a partial right
module over a partial operad $\P = (P = \coll E1,
\{\gamma_T :\u T \to E(n)\})$.
The constructions which we introduced for partial operads in
Section~\ref{rucicka}
carry over almost literally. For $S\leq T$, $S,T \in \T_n$,
put
\begin{equation}
\hat W(T) := \bigcup_{S\leq T} W_S(T) \subset M_E(T).
\end{equation}
As in the proof of Lemma~\ref{Pik}
we may show that the collection $\hat W =
\coll {\hat W}1$ defined by
\[
\hat W(n) := \coprod_{T\in \T_n} \hat W(T)
\]
is a topological submodule of the free right module $M\circ \F(E)$
generated
by the collection $M$
over the free
operad $\F = \F(E)$ (a strange
notation is justified by~\cite{dl}).

As in the case of partial operads, condition~(\ref{rest}) assures that
formula~(\ref{modular-extension})
defines the map $\nu_{S,T} : W_S(T) \to W_S(S)=: S_S$.
Let $\tilde \M = \coll{\tilde \M}1$ be given by
\begin{equation}
\tilde \M (n) := \hat W(n)/\sim
\end{equation}
with the relation $\sim$ identifying elements $\eta$ of $W_S(T)$ with
their
images $\nu_{S,T}(\eta) \in S_S \subset M_E(S)$.

In the following proposition, $W = \coll W1$ is the collection defined
in~(\ref{rest}), $U = \coll U1$ is the collection with $U(n) =
\u{T(n)}$,
$n\geq 1$
(compare~(\ref{21})), and $W \circ \F(U)$ is the free right
$\F(U)$-module
generated by the collection $W$.
}\end{odstavec}
\begin{proposition}
The collection $\tilde \M = \coll {\tilde \M}1$ is a topological right
module over
the operadic completion $\tilde \P$ of the partial operad $\P$.
There exists a natural epimorphism of topological right modules
\[
\delta: W\circ \F(U) \to \tilde \M.
\]
If the sets $W_S(T)$ are `combinatorially independent' in the sense
that
\begin{equation}
\label{Jituska-pusinka}
S',S'' \leq T,\ S'\not= S'' \Longrightarrow W_{S'}(T) \cap W_{S''}(T)
=
\emptyset,
\end{equation}
then the map $\delta$ is an isomorphism of sets.
\end{proposition}

The proof is essentially identical to the proof of
Proposition~\ref{casio}. We
call $\tilde M$ the {\em module completion\/} of the partial right
module
$\M$.

\section{Compactification of configuration spaces}
\label{sec5}

Let $V$ be an $m$-dimensional Riemannian manifold. Recall that $\Conf
nV$
denotes the space of configurations of $n$ distinct points in $V$.
This space
has a straightforward `framed' version
\[
\FConf nV := \{\vect x \times (f_1,\ldots,f_n);\
\vect x = (x_1,\ldots,x_n) \in \Conf nV,\ f_i \in F_{x_i}(V), 1\leq
i\leq n\},
\]
where $F(V)$ is the principal $\O(m)$-bundle of frames
on the manifold
$V$. Thus $\FConf nV$ is the space of configurations of $n$ distinct
points
of $V$, each decorated with
a frame. Another, fancier, way is to define $\FConf
nV$ as the pullback of the product bundle
$F(V)^{\times n} \to V^n$ under the
inclusion $\Conf nV \hookrightarrow V^n$.

If the tangent bundle of the manifold $V$ is trivial, then the
trivialization
defines an
isomorphism
$\FConf nV \cong \Conf nV
\times [\O(m)]^{\times n}$, which induces the inclusion
\begin{equation}
\label{www}
\Conf nV = \Conf nV \times [1\!\!1]^m \hookrightarrow \FConf nV
\mbox{ ($1\!\!1$ is the unit of $\O(m)$)}.
\end{equation}

We will define the partial right
module $\mu = (M=\coll M1,\{ \nu_T : \w T
\to W(n)\})$ (resp. $f\mu = (fM=\coll{f M}1,\{\nu_T :f\w T
\to fW(n)\})$) of (framed) virtual configurations of points
in the manifold $V$, over
the partial operad $\chi$ of virtual configurations
(resp.~over the partial $\O(m)$-operad $f\chi$ of framed
virtual configurations) of points in ${\bfR}^m$.
Let us start with the definition of $f\mu$.

The collection $fM$ is simply
$fM := \FConf{}V$. The definition of the subsets
$f\w T \subset fM_{fE}(T)$ is more
difficult. Observe first that, if $fM_{fE}(T) \not= \emptyset$, then
all vertices of
the tree $T$, except maybe the output one, have at least two input
edges;
we denote the set of all such $n$-trees by $\Tbine_n$. Let $\V :=
\vert(T)$, $\V = \{\ov\} \cup \V'$, where $\ov$ is the output vertex
of the
tree $T$. A typical element of $fM_{fE}(T)$ can be written as $\eta =
\{\lambda_v;\ v\in \V\}$, with
\begin{eqnarray}
\label{not2}
&\eta = \{\lambda_v;\ v\in \V\}, \lambda_{\ov} = \vect x \times \vect
f,\
\mbox{ where } \vect x = (x_1,\ldots,x_l) \in \Conf lV,&
\\
\nonumber
&\vect f = (f_1,\ldots,f_l),\ f_i \in F_{x_i}(V),\ l =
\inp(\ov),\mbox{ and}&
\\
\nonumber
&\lambda_u = (t_u,[\vect z_u], \vect g_w),\ t_u \in \bfR_{\geq 0},\
[\vect z_u] = [z^1_u,\ldots, z^{i_u}_u] \in \osfF_m (i_u), \mbox{ and
}&
\\
\nonumber
&\vect g_u = (g^1_u,\ldots, g^{i_u}_u) \in \O(m)^{\times i_u},\
i_u := \inp(u),
\mbox{ for } u \in \V'.&
\end{eqnarray}

For each vertex $u \in \V'$ there is an unique path in $T$ joining $u$
and
$\ov$ as in Figure~\ref{picture2} (with $u$ instead of $w_1$, $u_2$
instead of $w_2$, $\ldots$ , $\ov$ instead of $\ow$).
Put $x_u := x_{r_k}$ and $f_u :=
g^{r_{k-1}}_{u_{k-1}}\cdots g^{r_1}_u \cdot
f_{r_k}$.
The frame $f_u$ identifies
$\bfR^m$ with the tangent space $T_{x_u}( V)$ of
the manifold $V$ at the point
$x_u$, so we
may suppose that $[\vect z_u]$ is an element of
$\Conf{\inp(u)}{T_{x_u}(V)}/\Aff$. We may moreover suppose that
$\vect z_u = (z_u^1,\ldots,z_u^{i_u})\in \Conf{i_u}{T_{x_u}}$,
is {\em normalized} in the sense that
\[
\sum_{1\leq i\leq i_u} z^i_u = 0 \mbox{ and }
\sum_{1\leq i\leq i_u} |z^i_u|^2 = 1,
\]
where $|\!-\!|$ denotes the norm induced by the Riemannian metric.

For $1\leq i\leq n$ there exists in $T$ a unique
path from the $i$-th input to the output, as in Figure~\ref{picture2}
(with
$u_1$ instead of $w_1$, $\ldots$ ,$\ov$ instead of $\ow$.
Then put
\[
\varomega_i(\eta):= \exp_{x_{r_k}}(t_{u_{k-1}} \cdot
z_{u_{k-1}}^{r_{k-1}} +
\cdots + t_{u_1}\cdots t_{u_{k-1}} \cdot z^{r_1}_{u_1}),\
\varomega(\eta) := (\varomega_1(\eta),\ldots,\varomega_n(\eta))\in
V^n.
\]
It might seem strange, when we compare this formula to~(\ref{veprik}),
that
the coefficient at $z_{u_j}^{r_j}$ does not contain the product
$g^{r_{i-1}}_{w_{i-1}} \cdots g^{r_1}_{w_1}$. This is because this
expression
is already a part of the identification of ${\bfR}^m$ to the tangent
space
$T_{x_u}(V)$.

For $x\in V$, $v\in T_x(V)$ and $z:= \exp_x(v)$ define the `parallel
transport'
$\Phi_{x,z}:T_x(V) \to T_z(V)$ by
\[
\Phi_{x,z}(w) := \left.\frac{d}{dt}\right|_{t=0} \exp_x(v+tw),\ w\in
T_x(V),
\]
compare~\cite[(5.80)]{AS}. Define $\Phi(\vect f) :=
(\Phi(f_1),\ldots,\Phi(f_n))$, where $\Phi(f_i) :=
\Phi_{x_{r_k},\varomega_i(\eta)} (f_i) \in T_{\varomega_i(\eta)}(V)$.

Then $f\w T$ is the set of all $\eta \in fM_E(T)$ such that the points
$\varomega_1(\xi),\ldots, \varomega_n(\xi) \in V$ are distinct.
The structure map $\nu_T: f\w T \to fM(n)$ is defined as
$\nu_T(\eta) := \varomega(\eta) \times \Phi(\vect f)$, for $\eta \in
f\w T$.

As we already observed in~(\ref{www}),
if the tangent bundle of the manifold $V$ is trivial,
then the collection $M := \Conf{}V$ is a subcollection of the
collection
$fM = \FConf nV$ and, of course, $E$ is a subcollection of $fE$.
Thus we may put
$\w T := f\w T \cap M_E(T)$. We may moreover
suppose that the Riemannian metrics on
$V$ is induced by the trivialization. This means that the `parallel
transport'
$\Phi$ leaves the subcollection $\Conf{}V$ of $\FConf{}V$ invariant
and
$\nu_T$ restricts to a map (denoted by the
same symbol) $\nu_T :\w T \to W(n)$.

\begin{proposition}
\label{Ferdicek}
The object
\[
f\mu = (fM=\coll{f M}1,\{ \nu_T :f\w T \to fW(n)\}_{T\in \Tbine_n})
\]
is a partial right module over the partial operad $f\chi$ of framed
virtual configurations
and thus also over the partial suboperad $\chi \subset f\chi$.
It satisfies the
independence condition~(\ref{Jituska-pusinka}).

If $V$ is parallelizable, then the object
\[
\mu = (M=\coll{M}1,\{ \nu_T :\w T \to W(n)\}_{T\in \Tbine_n})
\]
is a partial $\chi$-submodule of the partial module $f\mu$.
\end{proposition}

There is an obvious framed version of
the Axelrod-Singer compactification $\CompConf nV$. Let
$\pi= (\pi_1,\ldots,\pi_n) :\CompConf nV \to V^n$ be the `blow down'
map, then put
\begin{equation}
\label{Andulka-pusinka}
\CompFConf nV := \{\xi \times (f_1,\ldots,f_n);\
\xi \in \CompConf nV,\ f_i \in F_{\pi_i(\xi)}(V), 1\leq i\leq n\}.
\end{equation}
As in~(\ref{www}), if $V$ is parallelizable, then $\CompConf nV$ is a
natural subspace of $\CompFConf nV$ for all $n\geq 1$.
Now we may
formulate the main theorem of this section.

\begin{theorem}
The module completion
$\widetilde {f\mu}$ of the partial module $f\mu$
coincides with
the framed version of the Axelrod-Singer
compactification $\CompFConf{}V$, $\CompFConf nV =
\widetilde {f\mu}(n)$ for any $n\geq 1$. This implies, among other
things,
that $\CompFConf{}V$ is a
natural right $f\sfF_m$-module.

If $V$ is parallelizable, then
the module completion
$\tilde \mu$ of the partial module $\mu$ is a natural right
topological
$\sfF_m$-submodule of $\widetilde {f\mu}$. It
coincides with
the Axelrod-Singer
compactification $\CompConf{}V$ of the moduli space of points in $V$,
$\CompConf nV =
\tilde \mu(n)$ for any $n\geq 1$.
\end{theorem}

As we have already observed in Proposition~\ref{Ferdicek}, we may also
consider $f\mu$ as a partial right module {\em over\/} $\chi$. One can
expext
that the module completion of $f\mu$ as a partial module {\em over\/}
$\chi$
will be smaller than the completion over $f\mu$. We leave to the
reader the
proof of the following proposition, see~(\ref{Andulka-pusinka})
for the notation.

\begin{proposition}
The module completion of $f\mu$ as a partial module over $\chi$
consists of
all elements $\xi \times (f_1,\ldots,f_n)\in FC_n(V)$ such that $f_i =
f_j$
whenever $\pi_i(\xi) =
\pi_j(\xi)$, $1\leq i,j \leq n$.
\end{proposition}

\section{Manifolds-with-corners and spectral sequences}
\label{sec6}

It is well-known that the spaces $\sfF_m(n)$ and $\CompConf nV$ are
manifolds with corners. Since $f\sfF_m(n) = \sfF_m(n) \times
\O(m)^{\times n}$ and, at least `locally',
$\CompFConf nV \cong \CompConf nV \times \O(m)^{\times n}$,
also the framed versions have structures of a manifold
with corners. We get immediately from~\cite[Proposition~1,
page~257]{Cerf}
the following proposition which says, roughly speaking, that the
compactifications discussed above are `truncations' of their open
parts.

\begin{proposition}
Each of the spaces $\sfF_m(n)$, $f\sfF_m(n)$, $\CompConf nV$ and
$\CompFConf nV$ is isomorphic to a closed submanifold (with corners)
of
its open part $\osfF_m(n)$, $f\osfF_m(n)$, $\Conf nV$ and
$\FConf nV$, respectively, obtained by removing a collar neigborhood
of the
boundary.
\end{proposition}

We need, however, a
deeper and more explicit understanding of these structures.
Recall that for a partial operad $\P$ and a tree $T\in \T_n$ we
introduced the
`open stratum' $R_T = U_T(T)$ and, similarly, for a partial right
module $M$
over $\P$ we have the `open strata' $S_T = W_T(T)$. Let $\T_n(p)$ be
the subset
of $\T_n$ consisting of trees with exactly $(p-1)$ vertices.

\begin{lemma}
\label{Katuska}
Let $\P = \chi$ or $f\chi$, then for each $T \in \T_n(p)$ there exists
a
`collar neigborhood' $\N(T)$ of the stratum $R_T$ in $\hat U(T)$,
isomorphic to
$R_T \times ({\bfR_{\geq 0}})^p$.

Similarly, for $M = \mu $ or $f\mu$, there exists a
`collar neigborhood' $\N(T)$ of the
stratum $S_T$ in $\hat W(T)$, isomorphic to
$S_T \times ({\bfR_{\geq 0}})^p$.
\end{lemma}

\noindent
{\bf Proof.\/}
Let us prove the lemma for $\P = \chi$, the proof of the remaining
three
cases is analogous.
Fix a tree $T
\in \Tbin_n$ and let $\xi \in Y(T)$ be as
in~(\ref{10}), i.e.
\begin{equation}
\label{spa}
\xi =\{ \kappa_w;\ w \in \W\},\ \kappa_w = (t_w,[\vect z_w])
\in[ {\bf R}_{\geq 0} \times \osfF_m(\inp(w))],\ t_{\ow} = 0,
\end{equation}
with $\W = \{\ow\} \cup \W' = \vert(T)$, where $\ow$ is the output
vertex of
$T$. We claim that for
any $\phi = \{[\vect z_w];\ w\in \W\} \in R_T =\osfF_m(T)$
there exists an
$\epsilon_\phi> 0$ such that, if $t_u < \epsilon_\phi$ for all $u \in
\W'$, then the element $\xi$ of~(\ref{spa}) lies in $U(T)$.
This follows from
the usual continuity argument and the observation that if all $t_u$'s
are `almost' zero, then certainly $\xi \in U(T)$.
Then the set $\N(T) = \{ \xi;\ t_u \leq
\epsilon_\phi,\ \phi \in \osfF_m(T)\}$ obviously has the required
property.
\qed

Consider the collection $\osfF_m = \coll{\osfF_m}1$
and the associated homology collection $\e_m
:= \coll{\e_m}1$ in the category of graded vector spaces given by
$\e_m(n)
:= H_*(\osfF_m(n))$. This collection is well-known to have
a natural structure of an operad. A
traditional way to show this fact is first to
observe that $\osfF_m(n)$ is
homotopically equivalent to $\Conf n{\bfR^m}$ (because $\osfF_m(n) =
\Conf
n{\bfR^m}/\Aff$ and the group $\Aff$ is
contractible) while the latter space is homotopically equivalent to
${\cal
D}_m(n)$, the $n$-th piece of the little disk operad.

The system of collar neighborhoods of Lemma~\ref{Katuska} however
defines this operad structure in a straigforward way.
Since we obviously have $H_*(\N(T)) = H_*(\osfF_m(T)) = \e_m(T)$,
the restriction $\gamma_T|_{\N(T)}: \N(T) \to U(n) = \osfF_m(n)$
induces, for each $T \in \Tbin_n$, a
map $\gamma^{\e}_T: \e_m(T) \to \e_m(n)$. We leave to the reader
the verification of the
following proposition.

\begin{proposition}
The system $\{\gamma^{\e}_T: \e_m(T) \to \e_m(n) \}$ defines an operad
structure on the collection $\e_m$. This structure coincides with the
structure induced by the little $m$-disks operad as explained above,
i.e.~the
operad $\e_m$ describes $n$-algebras in the sense of~\cite{GJ}.

Similarly, the partial operad $f\chi$
of framed virtual configurations induces an
operad structure on the collection $f\e_m := H_*(f\osfF_m)$, this
operad
describes $m$-dimensional analogs of Batalin-Vilkovisky algebras,
compare~\cite{bv}.
\end{proposition}

Analogous principle applies to the collections $f\m(V) := H_*(\FConf
{}V)$
(and $\m(V) := H_*(\Conf {}V)$ if $V$ is
parallelizable). As above we have the following
proposition.

\begin{proposition}
The partial $f\chi$-module $f\mu = (fM,\{ \nu_T :f\w T \to fW(n)\})$
of
virtual configurations of points in $V$
induces on the collection $f\m(V)$ a structure of a right module
over the operad $f\e_m$.

If $V$ is paralellizable, then there is an analogous right
$\e_m$-module
structure on the collection $\m(V)$ induced by
the partial $\chi$-module $\mu
=(M,\{ \nu_T :\w T \to W(n)\})$.
\end{proposition}

For an $n$-dimensional manifold with corners $M$, denote by $M[p]$ the
union
of the faces of $M$ with codimension $p$, and by $F_pM$ its closure
(sometimes called the {\em codimension $p$ skeleton\/}). Recall
Lemma~3.3 of~\cite{GJ} (but, since we do not assume $M$
to be compact, we must
work with the cohomology with compact supports).

\begin{lemma}
The filtration $F_pM$ induces a spectral sequence
with $E^1_{pq} = H_q(M[p])$ converging to
$H^{n-*}_{\rm comp}(M)$. The differential $d^1 : H_q(M[p])
\to H_q(M[p-1])$ is identified, by the Lefschetz duality, with the
boundary
map $\delta$ of the cohomology exact sequence of the triple
$(F_{p-1}M, F_pM,
F_{p+1}M)$.
\end{lemma}

The main theorem of this section uses the notion of the bar
construction over
an operad. This notion has already
became a standard one, we thus only briefly recall the
definition and refer the reader to~\cite{GK,GJ} for details.

Let $\Q$ be an operad in the category of graded vector spaces and
denote by $\susp \Q$ the suspension of the collection $\Q$,
i.e.~$\susp \Q =
\coll {(\susp \Q)}1$, where $(\susp \Q)(n) := \susp \Q(n)$ is the
ordinary
suspension of
the graded vector space $\Q(n)$.
The {\em bar construction on the operad $\Q$\/} is the differential
collection $\B(\Q) =
\coll {\B(\Q)}1$ with
\[
\B(\Q)(n) : = \bigoplus_{T\in \Tbin_n}(\susp Q)(T)
\]
and let the
differential $d_{\B} : \B(\Q)(n) \to\B(\Q)(n)$ of degree $-1$
defined as follows. Let
$T\in \Tbin_n$ and let
$e \in \edg(T)$ be an inner edge. If we denote by $T/e\in
\Tbin_n$ the tree obtained by collapsing the edge $e$, then the operad
composition on $\Q$ clearly defines a map $\delta_{T,T/e} :
(\susp\Q)(T) \to
(\susp\Q)(T/e)$. The differential is then given by
\[
d_{\B}(x) = \sum_{e\in \edg(T)} \pm \delta_{T,T/e}(x) ,\mbox{ for }
x\in (\susp Q)(T).
\]
The sign is a tricky part here. It is determined by demanding $d_\B$
to be
a degree $-1$ coderivation of the cofree
cooperad $\B(\Q)$. We do not need the exact formula for the sign here,
so we
just refer the reader to the above mentioned sources~\cite{GK,GJ}
for details.
Similarly, let $M$ be a right $\Q$-module. Let us define the {\em
bar resolution\/}
$\B(M,\Q) = \coll{\B(M,\Q)}1$ of the right $\Q$-module $M$ by
\[
\B(M,\Q)(n) = \bigoplus_{T\in \Tbine_n}(\susp M)_{(\susp Q)}(T)
\]
with the differential $d_{\B}: \B(M,\Q)(n)\to \B(M,\Q)(n)$
is defined analogically as the differential of the bar construction.
Compare also~\cite{GV,BJT}. Again, the bar resolution
$\B(M,\Q)$ can be shown to be a right comodule over the cooperad
$\B(\Q)$.
The case $M = \sfF_m$ of the following proposition was proven
in~\cite[Lemma~3.3]{GJ}.

\begin{theorem}
In the spectral sequence for the manifold with corners $M =
\sfF_m(n)$, the
term $(E^1,d^1)$ is naturally isomorphic to the $n$-th piece of the
bar construction $\B(\e_m)$ on the operad $\e_m$, $(E^1,d^1) \cong
(\B(\e_m)(n),d_\B)$.
Similarly, for $M = f\sfF_m(n)$, the first term is isomorphic to the
$n$-th
piece of
$\B(f\e_m)$.

For $M = \CompFConf nV$, the first term $(E^1,d^1)$ is isomorphic to
$(\B(f\m_m, f\e_m)(n),d_\B)$. If the manifold $V$ is paralellizable,
then, for $M =\CompConf nV$, $(E^1,d^1) \cong
(\B(\m_m,\e_m)(n),d_\B)$.
\end{theorem}

\noindent
{\bf Proof.\/}
Standard calculations show that the map $d^1 : H_q(M[p])
\to H_q(M[p-1])$ is induced by the inclusion $M[p] \subset M[p-1]$
which,
of course, does not exists in the literal sense. We must first thicken
$M[p]$, considered as a part of the boundary of $F_{p-1}M$, into a
collar
neighborhood and then move it a bit into the interior of this
neighborhood, which is a subset of $M[p-1]$.

We will describe this process in details for $M = \sfF_m(n)$
where the notation is
easiest, but all the remaining cases can be discussed in exactly the
same
way. We know that
the set $M[p]$ is the disjoint union of the open strata $R_T =
\osfF_m(T)$ over all trees $T \in \Tbin_n(p)$. Similarly,
$M[p-1] = \coprod \{ \osfF_m(S);\ S\in \Tbin_n(p-1)\}$. It is clear
from
the construction
that $R_T$ may intersect the closure of $R_S$ in $\sfF_m(n)$ if and
only if $S = T/e$, for some inner edge $e\in \edg(T)$.

Let $\phi = \{[\vect z_w];\ w\in \W = \vert(T)\} \in S_T$ and let
$w_0$ be
the input vertex of $e$. As in the proof of Lemma~\ref{Katuska}
there exists
$\epsilon_\phi > 0$ such that
\[
\{(t_w,[\vect z_w]);\ w\in W,\ t_w = 0 \mbox{ for $w \not= w_0$ },
t_{w_0} <
\epsilon_\phi \} \subset U_S(T).
\]
We may moreover suppose that $\epsilon_\phi$ depends continuously on
$\phi$. Then
\[
\tilde S_T :=
\{(t_w,[\vect z_w]);\ w\in W,\ t_w = 0 \mbox{ for $w \not= w_0$ },
t_{w_0} =
\mbox{$\frac12$}\epsilon_\phi \}.
\]
is an isomorphic copy of $S_T$ in $U_S(T)$. The corresponding
component of the differential $d^1$ is then induced by the composition
$S_T
\cong \tilde S_T \subset U_S(T)
\stackrel{\gamma_{S,T}}{\longrightarrow} S_{S} \subset
M[p-1]$. Our description of the operad structure then immediately
identifies
this map to $\delta_{T,T/e}$.\qed

\catcode`\@=11

\noindent
Mathematical institute of the Academy, 
\v Zitn\'a 25, 
115 67 Prague 1, 
The Czech Republic\hfill\break
e-mail: {\tt markl@mbox.cesnet.cz}

\end{document}